\pdfoutput=1

\documentclass[11pt]{article}

\usepackage{acl}
\usepackage{times}
\usepackage{latexsym}
\usepackage[T1]{fontenc}            
\usepackage[utf8]{inputenc}
\usepackage{microtype}              
\usepackage{inconsolata}            
\usepackage{graphicx}               
\usepackage{booktabs}               
\usepackage{comment}
\usepackage{float}
\usepackage{pgfplots}
\usepackage{amsmath}
\pgfplotsset{compat=1.18}
\title{CourtNav: Voice-Guided, Anchor-Accurate Navigation of Long Legal Documents in Courtrooms}

\author{
    Sai Khadloya \\
  \texttt{sai@adalat.ai} \\
  {\normalfont Adalat AI, India}
  \And
  Kush Juvekar\\
  \texttt{kush@adalat.ai} \\
  {\normalfont Adalat AI, India}
  \And
  Arghya Bhattacharya\\
  \texttt{arghya@adalat.ai} \\
  {\normalfont Adalat AI, India}
  \And
  Utkarsh Saxena \\
  \texttt{utkarsh@adalat.ai} \\
  {\normalfont Adalat AI, India}
}

\begin{document}
\maketitle

\begin{abstract}
Judicial work depends on close reading of long records, charge sheets, pleadings, annexures, orders, often spanning hundreds of pages. With limited staff support, exhaustive reading during hearings is impractical. We present CourtNav, a voice‑guided, anchor‑first navigator for legal PDFs that maps a judge’s spoken command (e.g., “go to paragraph 23”, “highlight the contradiction in the cross‑examination”) directly to a highlighted paragraph in seconds. CourtNav transcribes the command, classifies intent with a grammar‑first(Exact regex matching), LLM‑backed router classifying the queries using few shot examples, retrieves over a layout‑aware hybrid index, and auto‑scrolls the viewer to the cited span while highlighting it and close alternates. By design, the interface shows only grounded passages, never free text, keeping evidence verifiable and auditable. This need is acute in India, where judgments and cross‑examinations are notoriously long.In a pilot on representative charge sheets, pleadings, and orders, median time‑to‑relevance drops from 3–5 minutes (manual navigation) to 10–15 seconds; with quick visual verification included, 30–45 seconds. Under fixed time budgets, this navigation‑first design increases the breadth of the record actually consulted while preserving control and transparency.
\end{abstract}

\section{Introduction}
High‑volume courts routinely face long filings and crowded dockets (often dozens of matters per day) which leads to massive case delays \cite{agarwala2024mammoth}. Despite near‑universal digitization (e‑Courts) and access to case data at scale, the core interaction problem remains: \emph{how can a judge interrogate a voluminous record quickly and faithfully?}

Summaries aid orientation but can hide citations and miss pivotal passages, even retrieval‑augmented systems sometimes surface mis‑grounded references \cite{Survey2025Summarization, Stolfo2024Groundedness}. Adjudication prioritizes verifiability: decision‑makers must jump to the exact locus in the record and see it highlighted. We therefore target navigation, not paraphrase.

We present a voice-guided, \emph{anchor-first} navigator for long legal PDFs that converts a spoken command (e.g., ``go to paragraph~23'') into a highlighted paragraph within seconds. The system couples layout-aware indexing and anchor generation over scanned/structured PDFs, a constrained command grammar with LLM back-off for coverage, hybrid retrieval with de-duplication, and a viewer that auto-scrolls while preserving on-screen evidence. \textbf{Our primary contributions are:}
\begin{itemize}
    \setlength\itemsep{0em}
\setlength\parskip{0em}

  \item A court-facing system that prioritizes direct-to-paragraph, auditable navigation over free-form summarization. 
  \item A dataset and evaluation protocol for long-record navigation measuring time-to-relevance, strict-hit accuracy at anchor level, and end-to-end latency. 
  \item A pilot study on charge sheets, pleadings, and orders showing large reductions in time-to-relevance under fixed time budgets.
\end{itemize}

\section{Related Work}

\paragraph{Long-document QA and retrieval in law.}
Legal QA and retrieval have evolved from sentence-level factoid questions to long-form answers grounded in statutes and case law. Benchmark tasks span holding extraction (e.g., CaseHOLD \citep{zheng_casehold_2021}), case-retrieval datasets such as LeCaRD/LeCaRDv2 \citep{lecadr_sigir_2021,lecadv2_sigir_2024}, and broader evaluation suites like LegalBench \citep{legalbench_2023}. More recent resources target long-form QA (e.g., LLeQA, Legal-LFQA) \citep{louis_lleqa_2024,legal_lfqa_2024}. While these emphasize retrieval quality and reasoning, they operate at the document level, returning entire cases rather than pinpointed spans, and are not designed for judge-facing interaction loops.

\paragraph{Summarization for legal documents.}
Faithfulness remains a central challenge. Surveys and long-context datasets (e.g., CaseSumm) catalog hallucination modes and metric gaps \citep{basile_legal_summ_survey_2025,heddaya_casesumm_2024}. General summarization work similarly shows unsupported content in abstractive outputs \citep{maynez_faithful_sum_2020,fabbri_qafacteval_2022}. Summaries aid orientation but do not replace the need to \emph{jump to the exact place in the record}.

\paragraph{Evidence-first interfaces.}
Outside law, explainable QA resources require systems to surface supporting sentences (e.g., HotpotQA \cite{yang_etal_2018_hotpotqa}) and page-level localization for document images (DocVQA) \citep{Mathew2021DocVQA}, improving interpretability. However, most legal QA/summarization systems return text without a UI that \emph{enforces} verification.

\paragraph{}
Prior legal QA/summarization and DocVQA work does not focus on \emph{navigation} as we do: a voice-guided, anchor-first interface that maps spoken commands to highlighted paragraphs. Our system combines long-document indexing, hybrid retrieval, a domain-adapted query router, and a judge-facing viewer that \emph{enforces} verification. To the best of our knowledge, we are the first ones to attempt building such a system for the legal domain.

\section{System Overview}
\label{sec:system}

\begin{figure*}[t]
  \centering
  \includegraphics[width=0.95\textwidth, height=6cm, keepaspectratio]{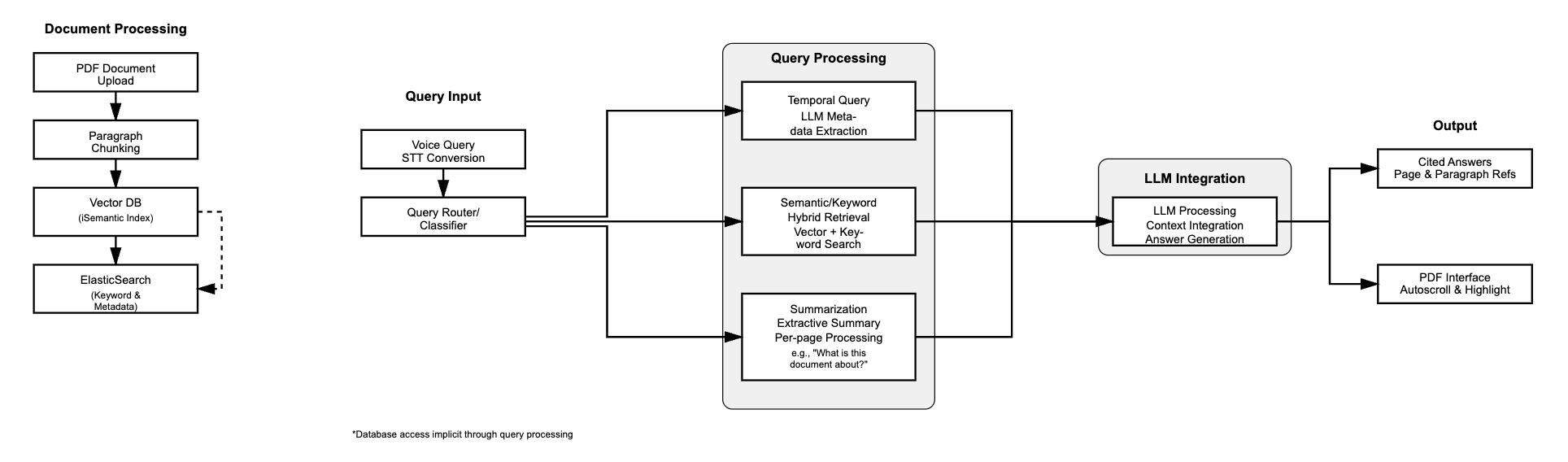}
  \caption{End-to-end flow. An uploaded PDF is parsed into layout-anchored spans and indexed (lexical + dense). Voice commands are transcribed on-prem and mapped to navigation actions. Retrieval produces candidate anchors, whose relevance to the queries is checked by the llm, while the viewer scrolls and highlights all the anchor which have substance related to the query}
  \label{fig:pipeline}
\end{figure*}

\subsection{Ingest and Layout-Aware Indexing}
Long records mix scanned pages, numbered paragraphs that reset per section, multi-column text, and tables that span pages. Pure text extraction loses the geometry needed for trustworthy highlights; vision-only pipelines are compute-heavy and brittle on low-quality scans. We therefore perform \emph{layout-aware parsing} that emits canonical spans with stable coordinates and IDs. \textit{Anchor (definition).} We treat every minimal displayable unit as an \emph{anchor} $\langle$\,page, bbox, span\_id, char\_range, type $\in $\{para, heading, table\_cell\}\,$\rangle$. Headings, paragraphs, and cross-page tables are extracted (e.g., with Docling) and normalized (hyphenation, numbering). We then build two complementary indices: a \emph{lexical} BM25 index for exact legal cues (sections, names, citations) and a \emph{windowed late-interaction} index for paraphrastic queries, produced over sliding windows to preserve local context \cite{Jha2024JinaColBERT}. For tables, we preserve grid structure (table\_id, row, col, rowspan/colspan) so cell-level anchors exist even when a table breaks across pages; we also store a light markdown/HTML rendering for downstream snippet previews \citep{docling2024,robertson2009bm25,khattab2020colbert,tabletransformer,layoutlmv3}.

\subsection{Query Interpretation and Routing}
Spoken requests cluster into three practical families: \emph{temporal} (“go to paragraph~23”), \emph{contextual} (“locate the contradiction in PW-2’s cross-examination about the call detail records”), and \emph{summarization} (“summarize the charges”). Latency and predictability are critical in court, so we use a \emph{grammar-first, LLM-backed} router. ASR text is first parsed by a compact command grammar that yields typed intents and slots (page/paragraph, statute, party, exhibit, or table region), if parsing fails or is ambiguous, a lightweight LLM back-off produces a structured action with confidence and a few disambiguating rewrites surfaced to the user. Summarization requests hit a precomputed extractive+abstractive synopsis, but responses still link back to anchors so users can inspect sources rather than accept paraphrase.

\subsection{Retrieval and Anchor Alignment}
A near hit is not enough; the system must land \emph{on} the paragraph (or cell). We perform hybrid retrieval across the lexical and late-interaction indices, interleave and deduplicate candidates by anchor overlap, then optionally re-rank a short list. Using the ingest-time anchor map, we deterministically map retrieved text offsets back to their anchors resolving OCR drift with tolerant matching and then command the viewer to smooth-scroll to the top anchor and \emph{highlight} all corroborating anchors. Table queries resolve to cell anchors via (table\_id, row, col) even across page breaks. If evidence is insufficient (low confidence or conflicting candidates), the UI offers a compact disambiguation list (keyboard/voice selectable) or withholds an answer. In all cases, every line of response is grounded in visible anchors rather than free text.

\subsection{Voice Pipeline}
Courtrooms are noisy, and users often code-switch. We run an on-premise streaming ASR pipeline (Whisper-based acoustic model with VAD gating and domain lexicon biasing for statutes, party names, and common legal terms)\citep{radford2022whisper, OpenAIWhisper2022} to generate partial transcripts quickly enough for responsive UI feedback . The Whisper model is fine-tuned on legal jargon and maintains an acceptable WER even in noisy ambient conditions through post-processing heuristics. The transcript, along with a ``confirm/cancel'' loop, gives the user opportunity to correct mishears and errors before any jump occurs. All audio is processed ephemerally, and nothing leaves the court network.

\subsection{Viewer and Interaction Design}
The UI is optimized for \emph{hands free, eyes busy} hearings. We extend a standard viewer (PDF.js) and design around three principles. \textbf{Speakable affordances:} every action a judge can perform via keyboard is also addressable through a short utterance with customizable shortcuts (next hit,'' previous section,'' toggle highlights''). \textbf{Anchored evidence:} the system never answers in free text without pointing to passages. All relevant anchors are highlighted with sentence-level backtracking to the anchor. \textbf{Low-drama navigation:} we prefer \emph{smooth scroll to anchor} rather than page jumps to preserve spatial memory. A breadcrumb trail records recent anchors and can be invoked to backtrack quickly. A compact \emph{evidence panel} lists retrieved snippets with page or paragraph badges, and clicking a badge or saying open two'' scrolls to that anchor. Keyboard shortcuts are supported for all operations so counsel can use the interface even if the microphone is muted. The layout avoids occluding the document, while transcripts and disambiguation chips collapse automatically after action, ensuring the judge’s visual context remains stable \citep{pdfjs}.

\subsection{Privacy and Deployability}
All components—ASR, router, retrieval, and viewer—run as independent services within the court’s infrastructure. No audio is stored, and logs capture only structured commands and anchor IDs for auditing. This design keeps the UI responsive under load while allowing each service to scale independently. The loose coupling also enables multiple judges to work concurrently without changing the user contract.

\section{Evaluation}
\label{sec:evaluation}

\subsection{Experimental Setup}
\label{sec:setup}
\textbf{Corpus and task construction.} To approximate day-of-hearing use, we curated long records that judges and counsel routinely handle: charge sheets (with annexures and lists), pleadings, orders, and reasoned judgments. Selection was stratified to cover (i) \emph{born-digital} and \emph{scanned} PDFs; (ii) table-heavy sections (accused/witness lists, seizure memos) and narrative sections; (iii) varied pagination/numbering schemes (paragraphs that reset, annexures, multi-column text). The final set has \textbf{15 documents} of \textbf{50--350 pages} each (avg.\ \textbf{100}). To elicit realistic queries, practising lawyers first skimmed each document as they would before a hearing and then authored speakable prompts in three families that reflect in-court needs: \emph{temporal} (explicit positions), \emph{contextual} (content descriptions), and \emph{summarization} (brief “what’s in the petition/charges” gists). Each query is paired with one or more \emph{gold anchor} paragraphs or table cells that must be annotated at anchor level and verified by a second lawyer, with disagreements adjudicated. The retrieval set comprises \textbf{600 contextual} and \textbf{50 summarization} queries. Temporal queries are generated directly from document numbering and appear across all documents.

\textbf{Participants and protocol.} For navigation trials, we recruit lawyers who did \emph{not} annotate the corresponding document. Each participant executes all queries for a document using two conditions: (i) a stock PDF reader (manual scroll and \emph{Find}), and (ii) \emph{CourtNav}. Conditions are counter-balanced across participants to mitigate order effects. Timing starts at query issuance (spoken or typed) and ends when the user lands on the gold anchor (temporal/contextual) or finishes a two-sentence synopsis with at least two paragraph-level citations.

\textbf{Baselines and measures.} The primary baseline is manual/search-based navigation with a stock PDF reader. Within our system we ablate retrieval modes: keyword-only, dense-only, hybrid, and our late-window+keyword variant. We report \emph{time-to-relevance (TTR)} in seconds and \emph{strict-hit F1} at paragraph (or table-cell) granularity, computed as mean~$\pm$~sd across participants and documents. For summarization, the baseline corresponds to the protocol above (producing a two-sentence gist with $\geq$2 citations using only the PDF reader), providing a practical comparator rather than full-document reading time.

\subsection{Results}
\label{sec:results}
Table~\ref{tab:results} presents time-to-relevance (TTR). The reader reduces TTR by half on Temporal commands ($t=13.3$, $p<10^{-7}$) and shortens Contextual queries from minutes to seconds ($t=58.6$, $p<10^{-12}$). For Summarization, we report only system time because manual reading scales with document length. The near-constant response time across query types stems from architectural choices: precomputed synopsis for summaries, direct anchor lookup for temporal spans, and sublinear vector search and fast elastic-search for retrieval \cite{malkov2018efficient}.

\begin{table}[t]
  \centering
  \begin{tabular}{lcc}
    \toprule
    \textbf{Query type} & \textbf{Baseline} & \textbf{Ours} \\
                        & (seconds) & (seconds) \\
    \midrule
    Temporal            & $10 \pm 2.0$ & $5 \pm 0.5$ \\
    Contextual          & $200 \pm 15.0$ & $6 \pm 1.0$ \\
    Summarization       & \textemdash\ & $6 \pm 1.2$ \\
    \bottomrule
  \end{tabular}
  \caption{Time-to-relevance (mean $\pm$ sd). Baseline is manual navigation with a stock PDF reader. “\textemdash” indicates no comparable baseline because manual reading depends on document length, and with our document length it scales to days}
  \label{tab:results}
\end{table}

\begin{figure}[t]
  \centering
  \begin{tikzpicture}
    \begin{axis}[
      ybar,
      bar width=18pt,
      width=0.9\linewidth,
      height=5.8cm,
      ymin=0, ymax=1,
      ylabel={Strict F1 Score},
      symbolic x coords={
        all-MiniLM-L6-v2,
        all-MPNet-base-v2,
        Keyword search,
        Hybrid search,
        Late-window+Keyword
      },
      xtick=data,
      x tick label style={rotate=45, anchor=east},
      nodes near coords,
      nodes near coords align={vertical},
      every node near coord/.append style={font=\small},
      ]
      \addplot coordinates {
        (all-MiniLM-L6-v2,0.43)
        (all-MPNet-base-v2,0.55)
        (Keyword search,0.70)
        (Hybrid search,0.85)
        (Late-window+Keyword,0.92)
      };
    \end{axis}
  \end{tikzpicture}
  \caption{Strict-hit F1 for different retrieval settings.}
  \label{fig:model_f1}
\end{figure}
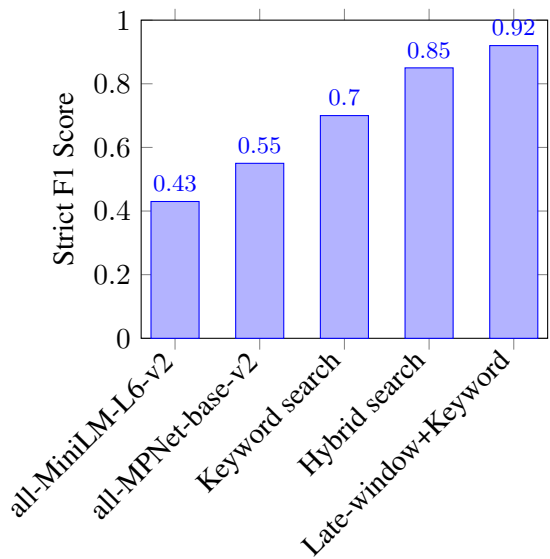

Retrieval choices significantly influence \emph{strict-hit F1} (Figure~\ref{fig:model_f1}). Keyword search performs well on statute or party mentions, dense-only aids paraphrase but misses exact citations, and a simple hybrid offers further improvement. However, our late-window+keyword variant achieves the best \emph{strict-hit F1} within the same latency budget.

\section{Conclusion}
We presented a voice-driven anchor-first reader that couples layout-aware indexing, hybrid retrieval, and a LLM-backed router to make long legal PDFs navigable in real time. In a pilot on charge sheets, pleadings, and orders, it cut time-to-relevance from several minutes to seconds (halved for \emph{temporal} jumps, orders-of-magnitude for \emph{contextual}) while preserving paragraph-level strict-hit accuracy and keeping every jump auditable. For Next steps, we will extend multilingual commands/ASR, and run field trials. We also release a long-form Indian legal retrieval dataset\footnote{\url{https://huggingface.co/datasets/adalat-ai/Indian-Legal-Retrieval-Generation}} which we plan to keep expanding, enabling Indian legal research.

\section*{Limitations}  
Our system currently supports documents up to 350~pages seamlessly, but as size increases, the responsiveness of the PDF.js reader declines. In future work, we plan to build a custom PDF viewer designed to operate smoothly with much larger documents. While the LLM-based query router shows strong accuracy in blind trials, absolute guarantees are impossible due to the stochastic nature of queries. RAG helps reduce hallucinations \cite{harvard_rag_2025, banerjee2024llms} , but does not fully eliminate them \cite{stanford_hai_2024}, even though we use a model adapted to strong instructions with explicit prompts to avoid ambiguous queries and to abstain when retrieved content is insufficient for a truthful answer. ASR errors are infrequent but non-negligible, and output varies with dialect or accent (especially given the wide range of accents in India). The system assumes English input, support for vernacular Indian languages remains future work on both the ASR side and document navigation side. A judge-in-the-loop feedback system is also missing, which will be essential for pilot testing and for developing stronger query classification models.

\section*{Ethical Considerations.}  Deploying AI in judicial settings raises ethical concerns. Generative models can reproduce biases present in training data, and their overconfidence may mislead users \cite{stanford_hai_2024}. We mitigate this by grounding answers in the document and by surfacing retrieved passages for verification.If no relevant retrievals exist, no answer is given, ensuring all responses remain strictly within the document. The system does not make substantive recommendations, it only navigates to requested text. User data is never sent to foreign APIs, is stored on Indian servers, and is deleted immediately upon user request. No data is used to train any models. We follow proper licensing, and all external software is open source under the Apache 2.0 License \cite{apache_license_2.0}. Our retrieval evaluation was fully transparent, but no benchmark covers every scenario due to the stochastic nature of information retrieval. We plan to improve incrementally by expanding the size of the dataset. 

\bibliographystyle{acl_natbib}
\bibliography{custom}

\appendix
\section{System User Interface}

\begin{figure*}[h]
    \centering
    \fbox{\includegraphics[width=1\textwidth]{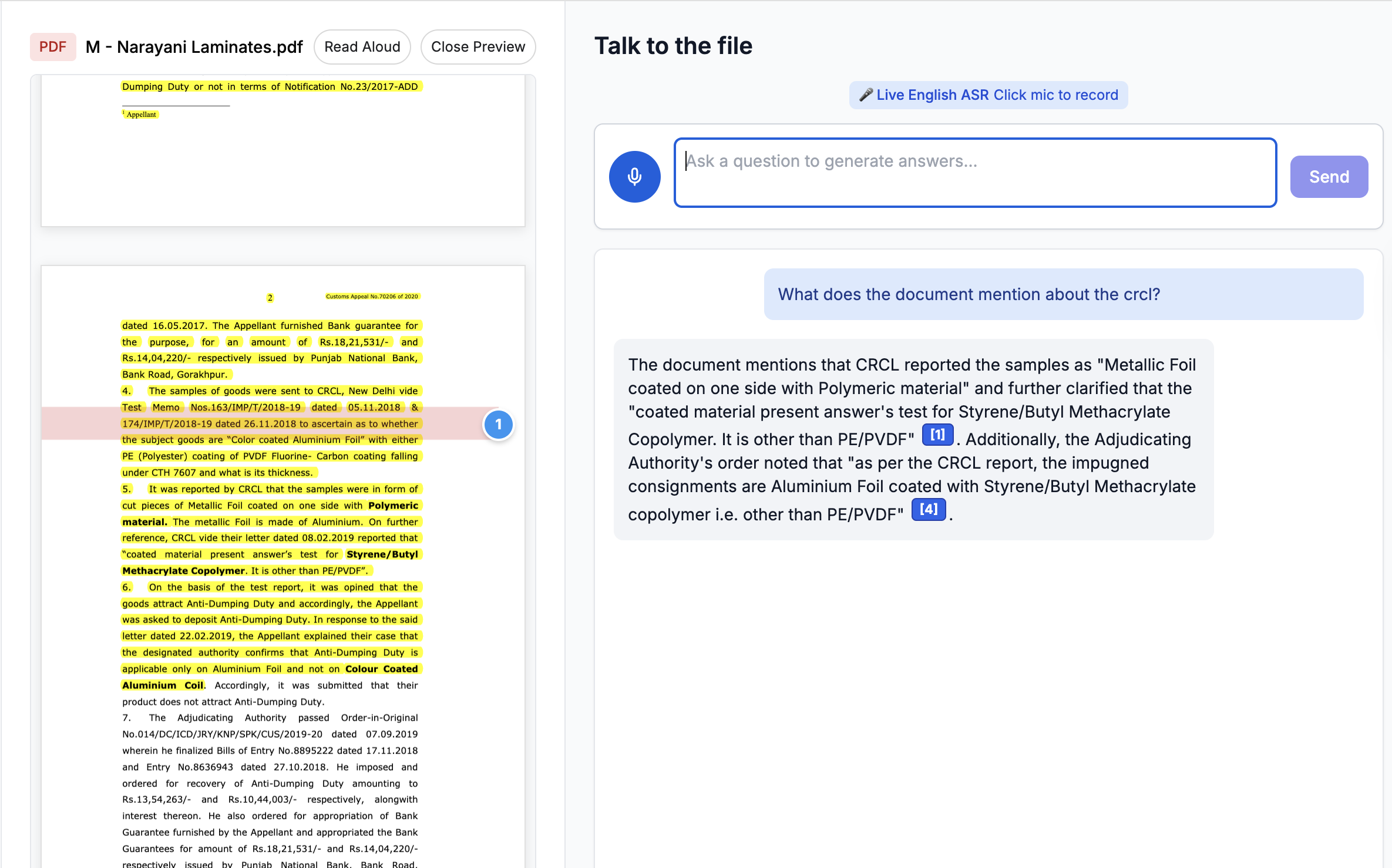}}
    \caption{User interface of the system showing the PDF viewer with document navigation capabilities and voice command interface.}
    \label{fig:system-ui}
\end{figure*}

The system interface demonstrates the core functionality described in Section~\ref{sec:system}, providing judges with direct document access through both voice and traditional input methods. The interface maintains the principle of anchored evidence display while supporting hands-free operation during hearings.

\section{Indexing Architecture Details}
\label{app:indexing}

\subsection{Elasticsearch Integration}
Our lexical indexing layer utilizes Elasticsearch 8.x as the primary engine for BM25-based keyword matching. The choice of Elasticsearch provides several advantages for legal document retrieval:

\begin{itemize}
\item \textbf{Legal-specific tokenization}: Custom analyzers handle legal citation formats, statute references, and party name patterns
\item \textbf{Field-specific boosting}: Paragraph headers, section titles, and table captions receive higher relevance weights
\item \textbf{Real-time indexing}: Supports incremental document addition during active court sessions
\end{itemize}

Index configuration includes custom mappings for legal document structure:

\begin{verbatim}
{
  "mappings": {
    "properties": {
      "content": {"type": "text"},
      "paragraph_id": {"type": "keyword"},
      "page_number": {"type": "integer"},
      "section_type": {"type": "keyword"},
      "bbox_coords": {"type": "object"}
    }
  }
}
\end{verbatim}

\subsection{Milvus Vector Database}
The dense retrieval component leverages Milvus 2.x for high-performance vector similarity search. Milvus provides:

\begin{itemize}
\item \textbf{Scalable vector storage}: Handles embedding collections for documents up to 350 pages efficiently
\item \textbf{GPU acceleration}: Supports CUDA-enabled similarity search for sub-second response times
\item \textbf{Index optimization}: Uses IVF\_FLAT indexing with 1024 clusters for optimal recall-latency trade-off
\item \textbf{Hybrid search support}: Enables metadata filtering combined with vector similarity
\end{itemize}

Vector collection schema:

\begin{verbatim}
collection_schema = {
    "chunk_id": DataType.VARCHAR,
    "embedding": DataType.FLOAT_VECTOR,
    "paragraph_anchor": DataType.VARCHAR,
    "document_id": DataType.VARCHAR,
    "page_range": DataType.VARCHAR
}
\end{verbatim}

\section{Late-Interaction Sliding Window Mechanism}
\label{app:late-interaction}

\subsection{Architecture Overview}
The late-interaction sliding window approach addresses two critical challenges in legal document retrieval: maintaining sufficient context for semantic understanding while preserving fine-grained anchor precision.

Traditional dense retrieval methods encode fixed-size chunks independently, potentially fragmenting legal arguments that span multiple paragraphs. Our windowed late-interaction mechanism operates as follows:

\begin{enumerate}
\item \textbf{Sliding window construction}: Generate overlapping windows of  paragraphs.
\item \textbf{Individual token encoding}: Each token in the window receives its own embedding vector
\item \textbf{Query-time interaction}: Compute similarity between query tokens and document tokens independently
\item \textbf{Maxpool aggregation}: Select maximum similarity scores across token pairs for final relevance scoring
\end{enumerate}

\subsection{Mathematical Formulation}
Given a query $Q = \{q_1, q_2, ..., q_m\}$ and document window $D = \{d_1, d_2, ..., d_n\}$, the late-interaction score is computed as:

$$\text{Score}(Q, D) = \sum_{i=1}^{m} \max_{j=1}^{n} \text{sim}(q_i, d_j)$$

Where $\text{sim}(\cdot, \cdot)$ represents cosine similarity between token embeddings. This formulation allows fine-grained matching while maintaining computational efficiency through maximum operations.

\section{Hybrid Search Implementation}
\label{app:hybrid}

The hybrid search combines Elasticsearch and Milvus results using a weighted scoring approach:

$$\text{Final\_Score} = \alpha \cdot \text{Keyword} + (1-\alpha) \cdot \text{Vector}$$

Where $\alpha = 0.7$ provides optimal balance for legal queries, emphasizing keyword matching while incorporating semantic similarity. Score normalization ensures comparable ranges across both retrieval methods.
\section{LLM Usage and Parameters for Reproducibility}
\label{app:llm}

All language understanding, summarization, and translation tasks within the pipeline were performed using the \textbf{Qwen3-Coder-30B-A3B-Instruct-FP8} model\footnote{\url{https://huggingface.co/Qwen/Qwen3-Coder-30B-A3B-Instruct-FP8}}, deployed via the \texttt{vLLM} inference engine for high-throughput serving.

The model operates in FP8 precision, enabling significantly reduced memory footprint and faster inference with negligible degradation in output quality. To ensure reproducibility, all experiments used the default \texttt{vLLM} sampling parameters unless otherwise stated.

\begin{itemize}
    \item \textbf{Model:} Qwen3-Coder-30B-A3B-Instruct-FP8
    \item \textbf{Serving Framework:} vLLM (GPU inference optimized)
    \item \textbf{Precision:} FP8 quantized weights
    \item \textbf{Max context length:} 8192 tokens
    \item \textbf{Default Sampling Parameters:}
    \begin{itemize}
        \item \texttt{temperature} = 0.7
        \item \texttt{top\_p} = 0.9
        \item \texttt{top\_k} = 50
        \item \texttt{repetition\_penalty} = 1.0
        \item \texttt{max\_tokens} = 2000
    \end{itemize}
    \item \textbf{Deployment:} Self-hosted GPU inference cluster
    \item \textbf{Integration:} Invoked via FastAPI microservice supporting both synchronous and streaming responses.
\end{itemize}

The combination of vLLM's optimized memory paging and Qwen's efficient A3B architecture provides low-latency, high-throughput inference suitable for real-time document understanding and generation workloads.

\section{Performance Optimization}
\label{app:optimization}

The document processing pipeline achieves real-time performance through:

\begin{itemize}
\item \textbf{Parallel processing}: Simultaneous embedding generation and Elasticsearch indexing 
\item \textbf{Connection pooling}: Persistent connections to both Elasticsearch and Milvus clusters
\item \textbf{Loose coupling}: ASR, Index stores and self hosted llms are loosely coupled and can scale independetly enabling a highly scalable and efficient architecture.
\end{itemize}

\end{document}